\documentclass{article}

\usepackage{arxiv}

\usepackage{hyperref}       
\usepackage{url}            
\usepackage{booktabs}       
\usepackage{amsfonts}       
\usepackage{nicefrac}       
\usepackage{microtype}      
\usepackage[utf8]{inputenc}
\usepackage[english]{babel}
\usepackage{amsmath,amsfonts,amssymb}
\usepackage{lmodern}
\usepackage[scaled]{helvet}

\usepackage{apacite}

\usepackage[T1]{fontenc}
\usepackage{lettrine} 

\usepackage{graphicx}

\newcommand{\cropped}[1]{%
  \includegraphics[width=0.8\linewidth]{#1_cropped.pdf}}

\newcommand{\croppedsmall}[1]{%
  \includegraphics[width=0.45\linewidth]{#1_cropped.pdf}}

\title{Epistemic Phase Transitions \\ in Mathematical Proofs}

\author{
  Scott Viteri\\
Department of Computer Science \\ 
Stanford University \\ Stanford, CA 94305 USA \\ \\
Social \& Decision Sciences \\
Carnegie Mellon University \\ 5000 Forbes Avenue \\
Pittsburgh, PA 15213 USA
   \And
 Simon DeDeo \\
Social \& Decision Sciences \\
Carnegie Mellon University \\ 5000 Forbes Avenue \\
Pittsburgh, PA 15213 USA \\ \\
Santa Fe Institute \\
1399 Hyde Park Road \\
Santa Fe, NM 87501 USA \\
  \texttt{sdedeo@andrew.cmu.edu}}

\begin{document}

\maketitle

\begin{abstract}
Mathematical proofs are both paradigms of certainty and some of the most explicitly-justified arguments that we have in the cultural record. Their very explicitness, however, leads to a paradox, because the probability of error grows exponentially as the argument expands. When a mathematician encounters a proof, how does she come to believe it? Here we show that, under a cognitively-plausible belief formation mechanism combining deductive and abductive reasoning, belief in mathematical arguments can undergo what we call an epistemic phase transition: a dramatic and rapidly-propagating jump from uncertainty to near-complete confidence at reasonable levels of claim-to-claim error rates. To show this, we analyze an unusual dataset of forty-eight machine-aided proofs from the formalized reasoning system Coq, including major theorems ranging from ancient to 21st Century mathematics, along with five hand-constructed cases including Euclid, Apollonius, Hernstein's {\it Topics in Algebra}, and Andrew Wiles's proof of Fermat's Last Theorem. Our results bear both on recent work in the history and philosophy of mathematics on how we understand proofs, and on a question, basic to cognitive science, of how we justify complex beliefs.
\end{abstract}


Mathematical proofs are often taken as a gold standard for certainty. How we come to justify and believe those proofs, however, is a different matter. In part because the things mathematicians believe are often remote from direct experience and what can be tested experimentally, the study of mathematical cognition has the potential to give new insights into the workings of the mind, and is a growing domain of research in cognitive science, psychology, and neuroscience~\cite{lakoff,goldstone,amalric2016origins}. At the same time, the explicit and idealized nature of mathematical proof can give us insight into a central cultural practice of the modern world, that of justification via abstract reason-giving~\cite{mercier2017enigma}.

This article focuses on the reception, rather than the discovery, of mathematical proofs. Discovery in mathematics is, famously, messy: a matter of heuristics, intuitions, and reasoning by analogy. When the mathematician shows her proof to colleagues, however, something quite different is at play. Heuristics, intuition, and analogy are all fallible~\cite{tall1981concept}. Understanding a successful proof, by contrast, is supposed to remove all doubt.

How? One view of proofs is that they are at heart specifications for, or summaries of, rule-based symbol manipulation. This picture began with \citeA{frege1882wissenschaftliche}, and became part of ``Hilbert's Program''~\cite{sep-hilbert-program}. Such a ``formalisms first'' picture appears most famously in~\cite{turing1937computable}. In this view, proofs specify a sequential, mechanical process of logical deductions that, operating according to purely syntactic rules, terminate with a formula corresponding to the thing to be proved. To validate that a proof is correct, one simply implements the program and checks for its proper termination.

The idea that proofs are (literally) programs is the basis of modern machine-aided proof systems~\cite{proofprogram}. Taken seriously as an account of human belief-formation, however, this model has many problems. Most relevant for our work, it implies that mathematical knowledge may be \emph{less} justified than ordinary beliefs. This is because, in an argument that goes back to \citeA{hume}, the possibility of errors in a sequential deduction compounds step by step~\cite{avigad2020reliability}. At an error probability of $10^{-3}$, for example, a proof with more than seven hundred steps is more likely to have failed than not.

Worse yet, mathematicians themselves often go out of their way to disclaim the precision required by the formalist picture. Henri Poincar\'{e} writes, for example, that he is ``absolutely incapable of adding without mistakes''~\cite{poincare1910mathematical}, a claim that, for the formalist, bears comparison to a neurosurgeon boasting of an unsteady hand. Even allowing for rhetorical exaggeration, the commonness of sentiments like Poincar\'{e}'s among practicing mathematicians points to additional epistemic processes for justification. Mathematicians describe, for example, posing questions to nature by the checking of cases~\cite{polya1954mathematics}, drawing on abductive or analogical principles~\cite{abduction_thesis,lakatos2015proofs}, and using intuition and idiom~\cite{bill}.

As part of resolving this paradox, we emphasize a key distinction. One the one hand, there is the question of how a proof establishes its claims; on the other, how we establish that it has established those claims. In the formalist picture, these are identical, but they need not be so in general. A proof may be a series of logical deductions, but we can come to understand how those deductions are correct by non-deductive methods. 

Our goal in this paper is to study how this might work. We present a simple model of how belief formation operates in judging deductive validity, and apply it to an unusual dataset of machine-aided proof networks. This provides a new account of how the deductive aspects of a mathematical proof interact with inductive and abductive justifications to produce certainty. 

We show that the epistemic relationship between claims in a mathematical proof has a network structure that enables what we refer to as an epistemic phase transition (EPT): informally, while the truth of any particular path of argument connecting two points decays exponentially in force, the number of distinct paths increases. Depending on the network structure, the number of distinct paths may itself increase exponentially, leading to a balance point where influence can propagate at arbitrary distance~\cite{stanley}.  Mathematical proofs have the structure necessary to make this possible. 

In the presence of bidirectional inference---\emph{i.e.}, both deductive and abductive reasoning---an EPT enables a proof to produce near-unity levels of certainty even in the presence of skepticism about the validity of any particular step. Deductive and abductive reasoning, as we show, must be well-balanced for this to happen. A relative overconfidence in one over the other can frustrate the effect, a phenomenon we refer to as the abductive paradox.

We present these results in three parts. We first introduce, and justify, a simple model of belief formation drawn from the cognitive science literature. We then describe the data sets we apply this model to. Finally, we present the results of the analysis. Technical details of the implementation can be found in the Methods Section.

\section{Cognitive Model}
\label{cm} 

Our model focuses on how the steps in a mathematical proof can combine to produce belief. To be clear at the outset, our model is not meant to capture the full range of things accomplished by a mathematical proof. Although belief is a core component---a proof that leaves room for doubt is not yet a proof---a successful proof does more than simply establish the truth of a theorem. 

Of any particular mathematical claim, for example, one might ask not just ``do I believe it?'', but also whether or not it is sensible, persuasive, generative, or elegant. One might consider a step in a proof in terms of its explanatory value rather than just its deductive support~\cite{wojtowicz2020probability}. One might ask how one step relates to others in the proof, and the role it might play in an overall strategy or proof schema~\cite{tall2012cognitive}. One might also try to judge the extent to which a claim corresponds to this or that part of one's own mental model, or to see how a claim relates to an internal image, object, or even affect built up in the course of reading~\cite{robert1998blending,selden2010affect}. It is only in rare circumstances that mathematicians approach a proof uncertain, in any real sense, of whether or not it contains an error.

At the same time, belief matters: a person does not fully understand a proof until they also see why, or how, the proof could establish the result with certainty. The bargain of our approach is that if we attend to how the steps of the proof can knit together to produce that certainty in the face of the Humean paradox, we will learn new things about the broader question of the psychology of proofs. ``Believing the axioms''~\cite{maddy1988believing} is an unusual turn of phrase for the practicing mathematician, but drawing on strategies from cognitive science, that talk in terms of degrees of belief, can reveal important patterns in the underlying representations.

Our model of belief formation is based on two core features of proofs. First, while proofs are usually presented in a linear narrative, most will refer back to claims made earlier. This turns a linear deduction into a network of interacting claims. A proof that combines two independent lines of reasoning at a particular point may be robust to counterexamples that invalidate one of those paths~\cite{robust}. As \citeA{von1956probabilistic} established in the case of faulty computer logic gates, multiple paths in a modular organization can overcome noise. 

Second, that reasoning on the basis of coherence, intuition, or analogy can also support evidentiary flow ``down'' from conclusions as well as ``up'', deductively, from axioms~\cite{maddy1988believing,corfield}. To take an extreme example, a proof that $1+1=2$~\cite{rw} can help establish the the consistency of the axioms and propositions that precede it, rather than resolving lingering doubts about elementary school arithmetic. A wide consensus on the truth of the Clay Institute's Millennium Prize Problems guides the mathematician's attempts to solve them, providing at least provisional support to supporting claims~\cite{villani}. 

P\'{o}lya's \emph{Patterns of Plausible Inference} is a famous argument for the importance of this downward direction that goes well beyond these celebrated cases to include more ordinary inferential moves that play a role in making sense of a proof. It corresponds to Peircean abduction~\cite{pei,eco1988sign}, an epistemic process that now plays a central role in the study of contemporary mathematical practice~\cite{zalamea}. One way to think about abduction in this Pericean sense is as a particular form of intuition---in this case, a non-deductive process in which one goes from a conclusion one believes to a felt sense of what would have to be the case to make the conclusion true. An abductive intuition may be related to other forms of mathematical intuition: for example, a felt sense that a conclusion must be true on the basis of an ``intuitive'' mental image not may lead, in turn, to an abductive intuition that this set of proof precursors (rather than, for example, another set) is the way that truth would have come about.

Abduction can, of course, be just as fallible as deduction, as demonstrated by long-standing gaps in proofs of famous theorems such as the Euler characteristic~\cite{formal_proofs}, where the intuitive truth of steps in the proof leads one to neglect flaws in more basic claims. 

\subsection{Network Structure}

To capture these non-deductive processes, we model mathematical belief-formation as the navigation of a network of claims where evidentiary support includes multiple, potentially bi-directional, pathways. Such networks are also the basic structures for coherence theories of belief formation in cognitive science~\cite{thagard1989explanatory,nont,jdm}. 

Here, each node of a proof network corresponds to a claim made as part of the proof. A claim might be ``$2n+1$ is odd'', for example, or ``the sum of angles in a triangle is always $180^\circ$''. In our machine proofs, the claims are more complicated formulae, expressed in in a type-theoretic logic. In neither case do we need to assert the existence of a privileged ``atomic scale''; in the machine-aided case, for example, a claim might be treated as a unit even if it could be expanded into more basic components, while human mathematical practice is to break proofs up into (for example) lemmas and subtheorems on the basis of pragmatic need.

Links between nodes are directional. Any particular claim is either an axiom, or depends upon other claims to establish its truth. Meanwhile, claims can be used, within the proof, to establish the truth of further claims. (For a concrete example, the reader might find it helpful to refer ahead to Fig.~\ref{graphs}.) This direction of inference we refer to as the ``deductive'' pathway. We also consider the possibility of inference in the other, abductive direction.

Abduction is never to be found in the proof itself. The proof itself is always a deductive argument that proceeds in a single direction, from axiom to consequent. Our model concerns the reception of that deductive argument by a reasoner. It is possible to use abductive methods to evaluate a deductive argument; knowingly increasing confidence in a deductive argument on partially abductive grounds is completely consistent with also believing that the argument itself establishes the truth deductively. We return to these issues in the discussion.

\subsection{Dynamical Belief-Formation} 

To specify a general model of belief formation on these networks, we invoke three requirements. First, we require that the degree of belief in any particular claim is (all other things being equal) conditional solely on its dependencies (the claims that support it in a deductive fashion), and those that depend on it in turn; we allow for differing strengths of these respectively deductive and abductive pathways. Second, following standard models in Bayesian cognition, we require this dependency be additive in log-space. Finally, we require that the model is otherwise unconstrained in the patterns of beliefs that can be formed. Taken together, these requirements lead to a minimal~\cite{billbirds} and unique ``maximum entropy'' model, corresponding to a probabilistic version of constraint satisfaction~\cite{tkavcik2013simplest,THAGARD19981}. Models of this form capture the fallible, real-world process of reasoning, not the normative process. The same network coherence perspective can model both (objectively) true and false beliefs; see, \emph{e.g.},~\citeA{rosenberg2006multiple}.  

Formally, we assume that each node in the proof tree can take on a (binary) truth value. We then model the (fallible) steps between claims in way that fixes the average error rate of a deductive step but leaves the system otherwise unconstrained~\cite{jaynes1957information}. For any particular error rate, some configurations (\emph{i.e.}, some assignments of ``true'' and ``false'' to the nodes) will be more likely than others. This corresponds to the Ising model,
\begin{equation}
P(\{t_i\})\propto \exp{\left(\beta\sum_{i,j} J_{ij}t_it_j\right)},
\label{ising}
\end{equation}
where $t_i$ is the truth value of claim $i$ (zero or one), the matrix $J_{ij}$ is non-zero when there is an evidentiary link between them (\emph{i.e.}, when $i$ invokes the truth of $j$ as part of its justification), and $\beta$ governs the reliability of the connection between $i$ and $j$; \emph{i.e.}, the extent to which the truth of $j$ given $i$ is believed to be correctly inferred. We measure the perceived truth as the time-average of the node; \emph{i.e.}, if the heuristic observer perceives the node to be true 70\% of the time, the overall degree of belief is $0.7$.

The statistical properties of this model can be simulated using the Metropolis-Hastings (MH) update rule. This can be thought of as a heuristic for belief update: a node shifts state (from believed-true to believed-false) depending on the state of its neighbours, and how the strength of that influence depends upon $\beta$. 

To aid the reader, we present the details of this algorithm using a worked example; we emphasise that the algorithm is chosen on the basis that it satisfies the three generic conditions described in the opening of this section, rather than its ability to capture the mechanistic details of a psychological process. The rule is phrased in a dynamical fashion, but nothing in our results depends on analyzing its dynamical properties; the purpose of the rule is that it enables us to approximate the stationary, equilibrium distribution given by Eq.~\ref{ising}, and there are a range of similar rules that lead to equivalent outcomes.

\begin{figure}[h!]
    \centering
\includegraphics[width=0.4\linewidth]{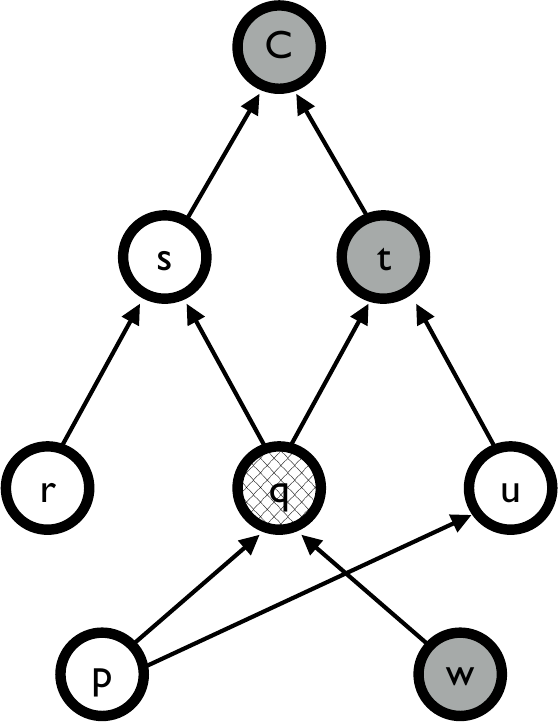}
\caption{A toy example of our belief formation model (see text). Nodes refer to claims, and arrows refer to directional dependencies; for example, claim $s$ depends on claims $r$ and $q$. At this stage in the belief-formation dynamic, three nodes (colored grey: $C$, the proof's final claim, and $t$ and $w$) are labelled ``false'', and node $q$ is under consideration. \label{demo_figure}}
\end{figure}
A very simple example of how the MH algorithm works in practice is provided by Fig.~\ref{demo_figure}. Here, we show a proof of the claim $C$, that depends on two claims ($s$ and $t$), that depend upon a sequence of precursors (for example, the truth of $s$ depends upon the truth of both $r$ and $q$), which themselves have precursors (for example, $p$ and $w$). We mark the current degree of belief by coloring the node grey for false, and leaving it white for true.

At the point in time shown in Fig.~\ref{demo_figure}, the person does not believe the proof claim $C$, but does believe a number of its parts. They are currently contemplating the step $q$ (marked by the cross-hatching). 

Three things come into play in the evaluation of claim $q$: the current state of belief in $q$, the state of belief in $q$'s direct precursors ($p$ and $w$), and the state of belief in the direct consequences of $q$ relevant to the proof ($s$ and $t$). The initial state of belief is that $q$ is true, that one of the precursors is true ($p$) and one is false ($w$), and that one of the consequences is true ($s$) and one is false ($t$).

The MH prescription provides a rule for when to update the truth value of $q$. The rule is in some cases deterministic: in particular, if changing the truth value of $q$ (in this case, from true to false) brings $q$ into alignment with a majority of the nodes it's connected to, then the prescription is to always make the switch. Otherwise, the rule is deterministic: if such a change would not have that effect (either because it is already reflective of the majority opinion, or because opinion is equally divided) then the change should be made with probability $P$. That probability depends on both $\beta$ (the strength of the influence of the neighbours of the node) and $D$, defined as the number of nodes that currently agree with $q$ minus the number of nodes that would agree with $q$ if the value were flipped; $P$ is equal to $\exp{(-2\beta D)}/(1+\exp{(-2\beta D)})$. In this case, $D$ is zero, and so $P$ is one-half.

MH is the simplest sort of update rule, but can be shown that a range of possible rules are consistent with the underlying stationary distribution Eq.~\ref{ising}; for example, so-called Glauber dynamics~\cite{Martinelli1999} allow for updates to be stochastic even in the case where changing the truth value would bring the node into alignment with the majority of its neighbours. A more sophisticated version of the update rule would allow for degrees of belief; in the simple version used in this paper, one can consider the degree of belief in a node to be equal to the time-averaged version over the course of the process.

We emphasise that the update rule is intended to capture the heuristics of examination and not the outcome of a deductively correct process. First, and most obviously, it includes an abductive direction, with evidence for a logical consequence of a belief allowed to play a role.

Second, it allows evidence to ``accumulate'' in the deductive direction. Consider, for example, the case of node $q$. As a matter of deductive logic, the only case in which one can always draw a conclusion is when the argument is valid, \emph{i.e.}, both precursors are true, $p$ and $w$. By contrast, the MH model allows evidence for $q$ to arrive in pieces, where believing in $p$ makes one more likely to believe in $q$ regardless of the state of belief in $w$.

A natural extension to the MH rule is to account for the differential impact of deductive dependencies and abductive implications, \emph{i.e.}, a different value of $\beta$ depending on whether the coupling is from $i$ to $j$ or from $j$ to $i$. This leads to the asymmetric Ising model, where the effect of $A$ on $B$ (all other things being equal) may not equal the effect of $B$ on $A$, used in  studies of updating in game-theoretic contexts~\cite{galam2010ising}. We write the strength of a dependence as $\beta_\mathrm{dep}$, and the strength of an implication is $\beta_\mathrm{imp}$ (\emph{i.e.}, the extent to which belief in a claim derivable from $A$ abductively increases confidence in $A$).

Consider, for example, what happens when evaluating claim $t$ in Fig.~\ref{demo_figure}, under the assumption that $q$ has been updated to true. In the simple case, the MH rule deterministically updates $t$ to be true: the fact that a consequence of $t$ (claim $C$) is false is outweighed by the ``evidence'' for $t$ contained in its logical precursors. However, by increasing $\beta_\mathrm{imp}$ relative to $\beta_\mathrm{dep}$, one can make disbelief in $C$ overwhelm the deductive evidence, making it possible to $t$ to remain ``considered false''.

(In passing, we note that is possible to consider scaling $\beta_\mathrm{dep}$ with $n_\mathrm{dep}$, the number of dependencies. This would capture the idea that increasing the number of dependencies for any particular node should not affect the relative impact of that node's abductive pathway. However, because the distribution of dependencies is not very wide, however, this amounts in practice to a simple rescaling and does not significantly affect our results.)


Having specified the update rule, all that remains is to specify the initial conditions. We begin our simulations with a (weak) bias in favor of truth, here a weakly charitable predisposition for the reader to consider the proof more likely to be true than false at the level $p_{\mathrm{prior}}$ before considering evidentiary links. To determine the final state of the network, we iterate the rule for ten times the total number of nodes in the network, corresponding to (on average) looking over each step of the proof ten times. This is sufficient to reach the equilibrium distribution corresponding to Eq.~\ref{ising}. Empirically, there are no significant differences to alterations in the details of the MH rule, such as different methods for choosing which node to examine next; this is as expected---as long as the rule obeys what is known as the detailed balance conditions, we will recover Eq.~\ref{ising}.

\subsection{Three Consequences of Belief Dynamics}

Despite its simplicity, this model makes (as we shall show empirically, in the next section) three predictions directly relevant to how we perceive and evaluate proofs. First that, under certain conditions, the very things that lead to the reliability problem can become a virtue. The physical analogues of our model are known to undergo phase transitions, where small changes in a control parameter (in our model, the value of $\beta$) can lead to discontinuous shifts in global properties. Here, in the cognitive domain, the control parameter corresponds to the local degree of dependency (\emph{i.e.}, how good the reasoner is, or thinks he is, about drawing the correct conclusion from the premises), while the global property is the average degree of belief in a claim.

The existence of such transitions is sensitive to network structure: they can not happen, for example, for a linear network, nor indeed for any tree-like network of finite ramification~\cite{gefen1984phase}. However, when the topological demands \emph{are} met, justification becomes easier, not harder, as the network size, $N$, increases, with---depending on the structure of the network---a sharp transition to total deductive certainty in the limit of large $N$. 

As mentioned above, this is due to the emergence of multiple paths between any two claims. An example of this appears in Fig.~\ref{demo_figure}, where belief in the axiom $p$ influences the belief in the claim $t$ (and vice versa, abductively) via two pathways: from $p$ to $q$ to $t$, and from $p$ to $u$ to $t$. 

To see how multiple paths combine in the MH model, imagine fixing the value of $t$ to be true. This will make it more likely that $q$ will update to true (or remain true) when it comes to be considered. This leads to a second-order effect, of $q$ updating to true influencing $p$ updating to true. Fixing $t$ to true also makes it more likely that $u$ will update to true, leading to an additional, otherwise independent, effect of $t$ on $p$. Each of these second-order effects is weaker than a direct link, since it requires that the outcome of two random processes comes out in the more-likely direction: $q$ flipping because of $t$, then $p$ flipping because of $q$, for example. At the same time, however, the two nodes can have independent effects on $p$. (Second-order effects can also emerge from a combination of abductive and deductive steps; for example, $s$ can influence $t$ both via belief in $q$, and via belief in $C$.)

What this line of reasoning reveals is that the effect of one part of a proof on another part of the proof depends upon the network topology, and in particular the number of distinct paths between the two parts. In general, the power of the effect along any particular pathway decays exponentially in the number of steps; this is another way of saying that error compounds exponentially along any particular path, as it does in any linear chain. At first glance, this suggests that, as a proof becomes larger, it ``falls apart'' in the expected Humean fashion: the truth of one claim has vanishingly little impact on something in an entirely different part of the system.

On the other hand, the number of distinct paths between points can, depending on network structure, grow exponentially~\cite{dedeo2012dynamics}: the further away you are from a particular claim in the network, the more non-overlapping pathways exist to get there. At a critical point, the exponential decay is balanced by the exponential growth, and influence can propagate undiminished across the entire network~\cite{stanley}. Informally, the EPT corresponds to the perception of separate lines of evidence, or reasoning, all pointing in the same direction. Our first result in this work is that mathematical theorems have the particular structure necessary for an EPT to take place, and for conclusions to become stronger than any particular step.

EPTs are a double-edged sword, however, because disbelief can propagate just as easily as truth. A second prediction of the model is that this difficulty---the explosive spread of skepticism---can be ameliorated when the proof is made of modules: groups of claims that are significantly more tightly linked to each other than to the rest of the network. Imagine, for example, that node $r$ in Fig.~\ref{demo_figure} was no longer taken as an axiom, but was the conclusion of a logical inference with nodes $r^\prime$, $r^{\prime\prime}$, and so on that led to $r$ as a conclusion. The connections between these $r$-nodes might well be dense enough to trigger an EPT. However, their connection to the rest of the proof in Fig.~\ref{demo_figure} is mediated by the single node $r$. Such a situation would correspond to the presence of (at least) two modules: the $r$-nodes correspond to one module, and the remainder of the network to (at least) one other module.

When modular structure is present, the certainty of any claim within a cluster is reasonably isolated from the failure of nodes outside that cluster. In the case of our example, disbelief in all of the non-$r$ nodes would not necessarily be a hinderance to forming a belief about the truth of the $r$-node module; even if disbelief in $s$, for example, led one to doubt $r$, this would not necessarily propagate further inward to $r^\prime$, $r^{\prime\prime}$, and so on, for the same reason that disbelief in $C$ is insufficient to prevent belief revision in favor of the theorem. 

Effects of this sort means that a reasoner can come to believe sub-parts of the overall network without needing to resolve the entire system at once. Such a mechanism has been hypothesised to exist in the case of mathematical proofs~\cite{avigad2018modularity}.

Modules can be identified by standard clustering algorithms such as Girvan-Newman~\cite{newman2004finding}. The subsequent robustness can be tested by comparing the relative difficulty of forming a belief within a module compared to that of forming a belief in an arbitary collection of nodes. This is operationalized as $\Delta\mathcal{L}_1$, the relative log-likelihood per node of the two cases, computed using Eq.~\ref{ising}; see Appendix for further details.

A third prediction of the model concerns the balance of deductive and abductive reasoning. Informally, one would imagine that increasing confidence in either process would aid the overall confidence in the proof: at a given level of deductive confidence (say), I can only become more certain by increasing confidence in my abductive intuitions.

This turns out not to be the case: for a fixed level of deductive confidence, increasing abductive confidence can lead to \emph{lower} levels of certainty, and similarly in reverse. This is because, at a critical point, abduction can come to dominate deduction completely, leading to solely downward propagation and cutting off the proliferation of paths necessary for the EPT. This destroys key topological features necessary for a phase transition: there are now fewer paths of influence between nodes and, for example, theorem ``siblings'' can no longer re-enforce each other. 

We refer to this as the abductive paradox. Informally, the proof becomes dependent solely on the mathematician's belief in the conclusion: doubts propagate downwards, and even the best axioms are powerless to overcome them. 




\section{Data}

In order to determine if real-world mathematical theorems have the necessary properties to trigger the epistemic effects described in the previous section, we analyze two datasets. First, forty-eight machine-assisted proofs constructed by mathematicians with the aid of the formal verification system Coq~\cite{plugin}, ranging from the Pythagorean Theorem to the Four-Color Theorem; see Table~\ref{alpha_list}.

Proofs in a formal verification system like Coq are constructed by a mathematician who helps the machine find the necessary epistemic structure by intelligent use of automated heuristics. The output of this process is a program equivalent to a proof. For each of the programs, we extract the abstract syntax trees representing the underlying deductions, and then identify of equivalent claims which turns these trees into directed acyclic graphs. The proofs themselves are interpretable, if exceedingly pendantic; see the Appendix for a proof that the number four is even. They provide us with an (ultra) high-resolution picture of the relationship between claims in mathematical reasoning; as we discuss further below, our goal in using Coq is not to answer questions about the reliability of machine inference, but to help in the project of understanding human reason.

The Coq dataset is supplemented with five ``human'' proofs: the original texts of Euclid's \emph{Elements} and Apollonius' \emph{Conics}, both of which mark explicit dependencies, and three hand-coded networks built from contemporary mathematical practice. These are (1) Andrew Wiles' 1995 proof of Fermat's Last Theorem~\cite{wiles1995modular}, (2) \citeA{orlik2015jordan}, a proof in group theory chosen as a recent example from the \emph{Journal of the American Mathematical Society}, and (3) the full text of Israel Nathan Herstein's \emph{Topics in Algebra}~\cite{herstein1975topics}. ``Herstein'' is widely-used in upper-level undergraduate and graduate-level mathematics courses that proves an interconnected network of theorems in abstract algebra; it appears on 1,412 syllabi in the Open Syllabus Project.\footnote{\url{https://opensyllabus.org/result/title?id=53034256171694}} Out of an abundance of caution, we read and constructed the human proof networks by hand, a somewhat laborious process, rather than relying on automated NLP tools. Although human networks are orders of magnitude smaller than those supplemented with the fine-grained deductions constructed by aid of a machine, the comparison allows us to find, and confirm, the similarities between the two.

The simultaneous use of both human and machine-aided proofs allows us to bracket, or bound, the complications of real-world mathematical reasoning. Our human proof networks, by necessity, connect together claims that the mathematician has taken the trouble to name or number. Each one of these claims, of course, usually contains a complex sequence of finer-grained argumentation that could, if desired, and given sufficient time, be broken down into sub-parts which form their own network, but were not. Machine proofs, by contrast, name and number the steps of mathematical reasoning at far higher resolution; comparing the number of nodes in the human and machine version of Euclid's \emph{Elements} for example, suggests that the relative resolution is on the order of 400-to-1: for every numbered claim in Euclid, the machine version has nearly 400 steps.

Actual human reasoning occurs at a resolution somewhere in between these two scales. Consider, for example, Theorem 5.3.1 of Herstein; it cites two prior steps, but reader who sees only that the theorem depends on those two steps (Lemma 3.9.6 and Theorem 3.5.1) is missing a great deal of the argument's structure. The theorem assembles together a whole sequence of additional claims, including a proof by induction and two un-named steps marked as deserving further unfolding by the student; all of these steps are ones that are plausibly shared with other proofs, and understanding the proofs in the chapter is in part understanding these more micro-level moves. 

At the same time as we recognize that the named-dependency structure is too coarse-grained to capture everything going on, it is also the case that no student is expected to dig as deeply, or as rigorously, as a machine-aided proof. The machine proof is expected to be much more fine-grained than human practice. Our use of Coq proofs is thus, partly, opportunistic; properties that the machine and the human scale have in common are at least plausibly something shared with the intermediate scales that human mathematicians are expected to work at.

The use of both kinds of proofs also provides an important check on our claims of the epistemic status of mathematical knowledge. Human mathematicians may have introspective access to the ways in which they come to believe something, but there is no guarantee that they match the actual reasoning process itself. Machines, through expansions that are orders of magnitude larger than the self-reported steps in the corresponding human case, make visible what is idealised and implicit in human communication. Despite the vast technological gulf between them, the machine-aided proofs of the twenty-first century share, as we shall see, the same basic epistemic properties as Euclid's more than 2500 years prior.

\section{Results}

We present our results in three sections. First, the topological properties of the proof networks; second, the emergence of epistemic phase transitions and the existence of modular firewalls; and finally, the abductive paradox.

\subsection{Network Structure}
\label{net}

\begin{figure}[h!]
    \centering
    \begin{tabular}{cc}
    \croppedsmall{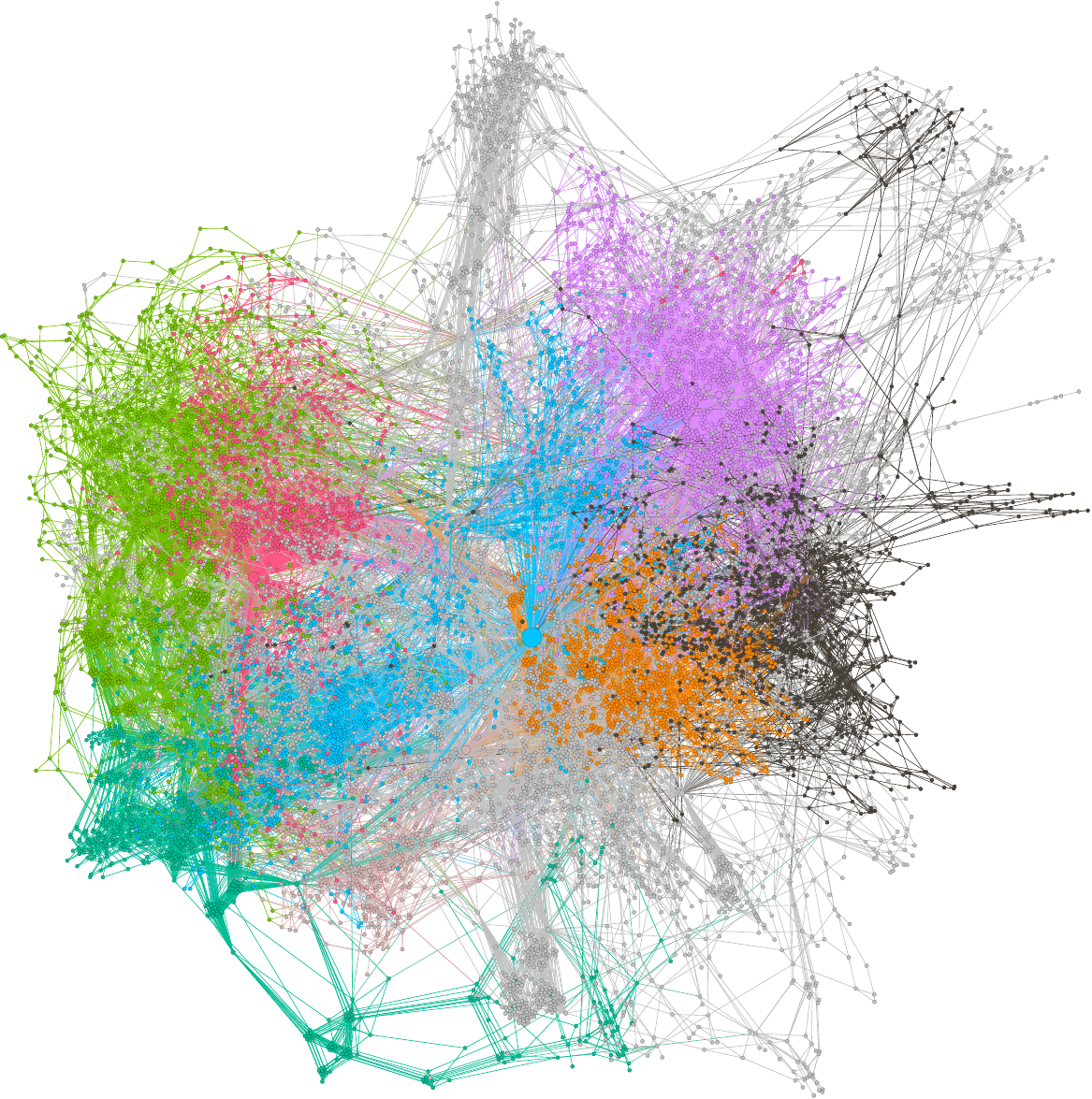} & \croppedsmall{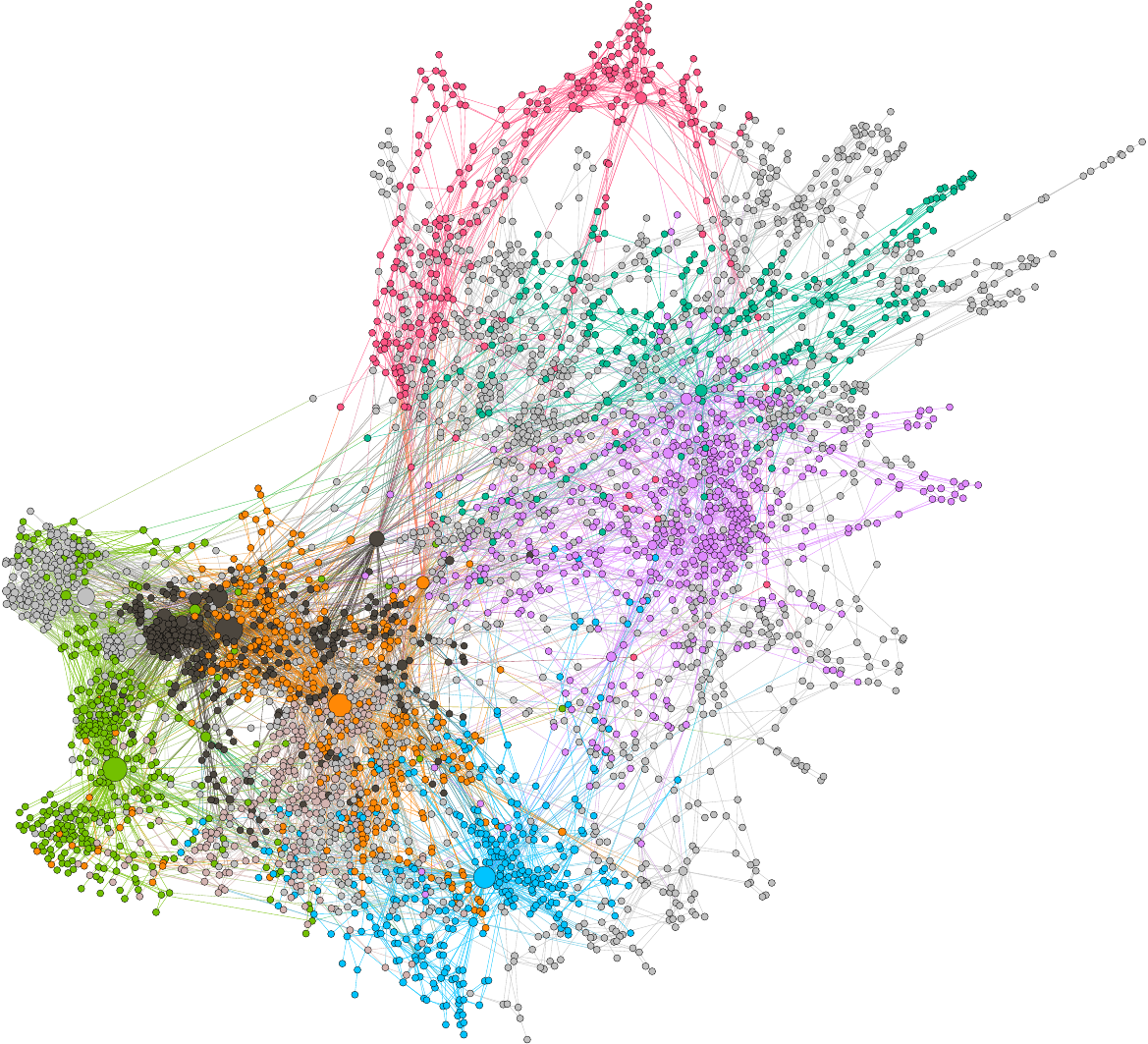} \\
    \croppedsmall{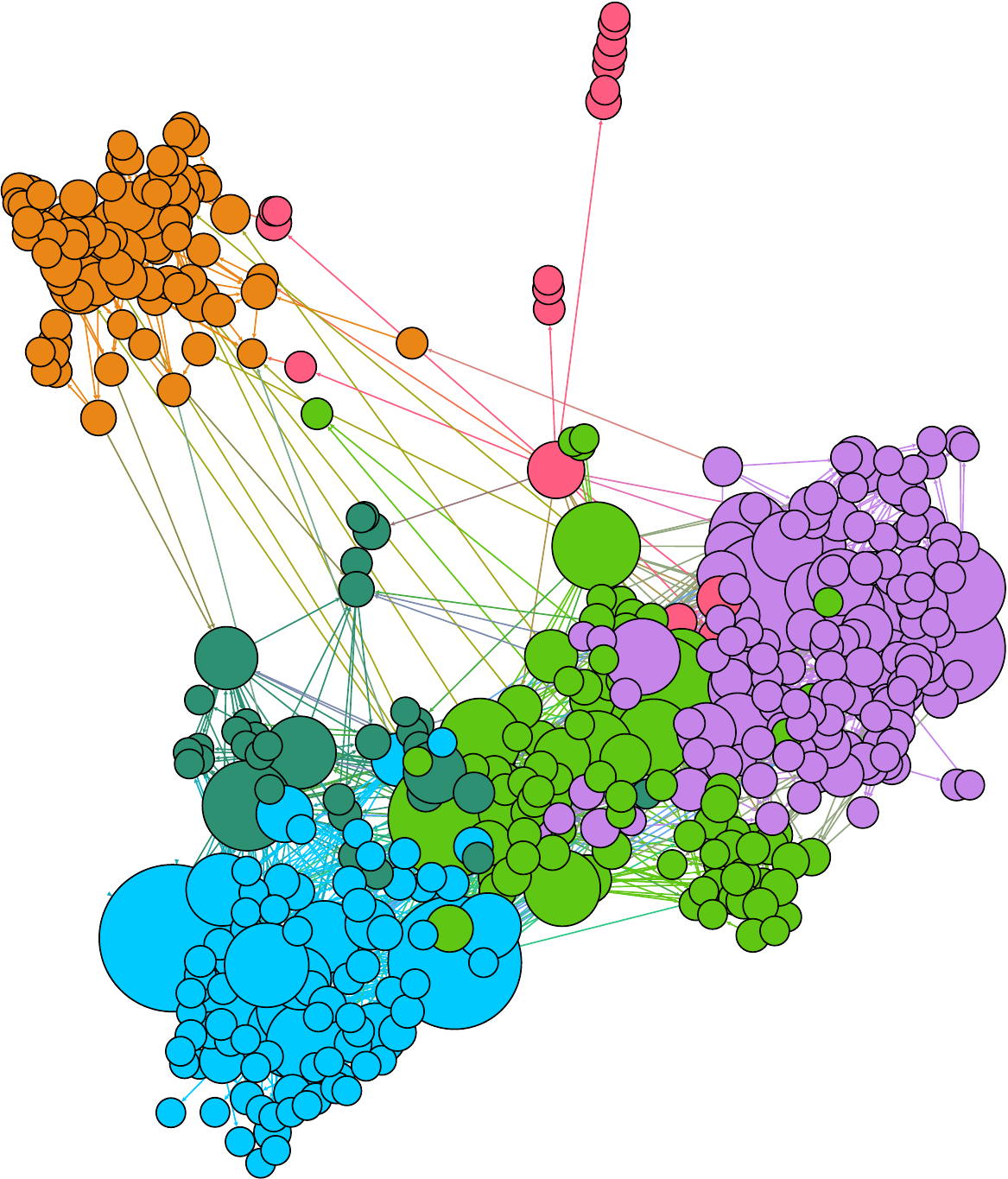} & \croppedsmall{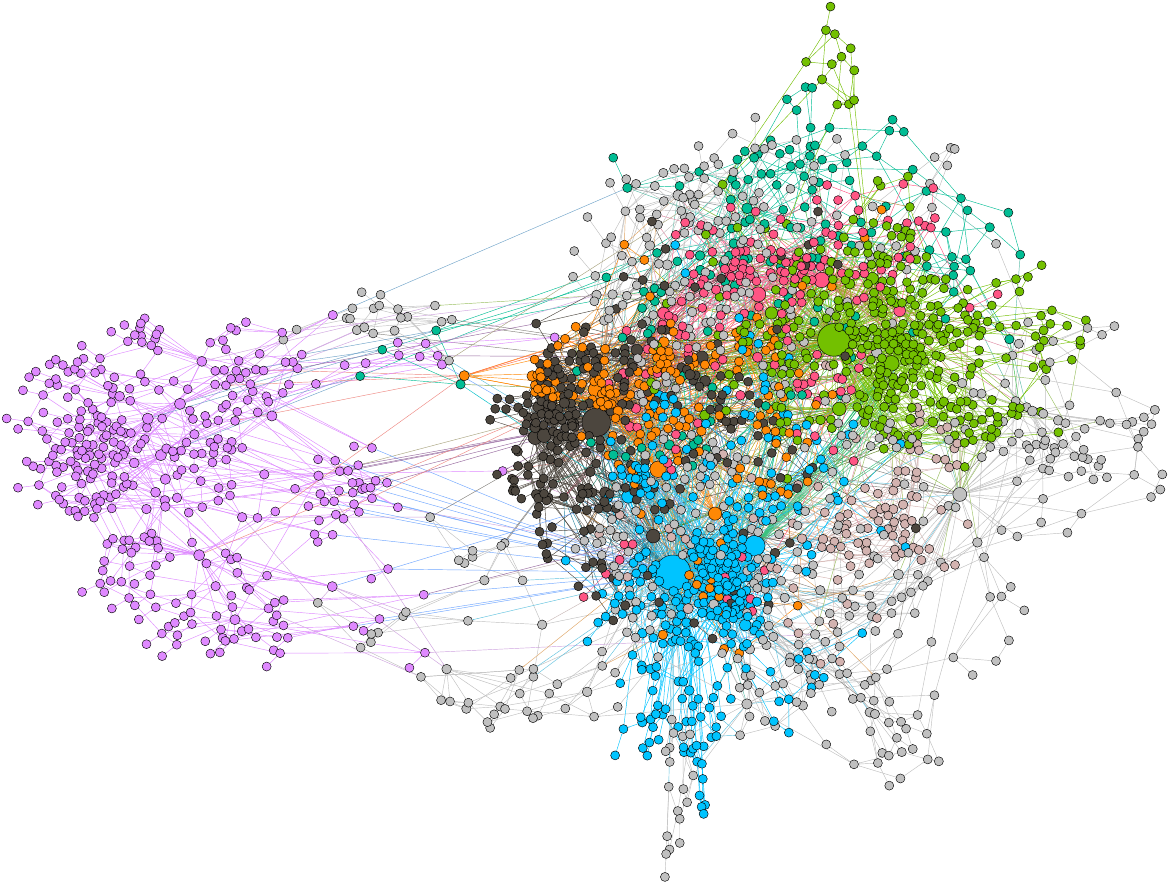}
    \end{tabular}
    \caption{Implication structure for four proof networks in our database. Clockwise from top left: the Four Color Theorem (Coq), the uncountability of the Reals (Coq), G\"{o}del's First Incompleteness Theorem (Coq), and Euclid's \emph{Elements} (original Greek Text). We color the top clusters by membership, and size nodes according to out-degree (\emph{i.e.}, the number of nodes that have that node as a deductive pre-requisite). Both human and machine-aided proofs are characterized by high levels of modularity, and a heavy-tailed distribution of out-degree.}
    \label{graphs}
\end{figure}

\begin{figure}[h!]
    \centering
    \cropped{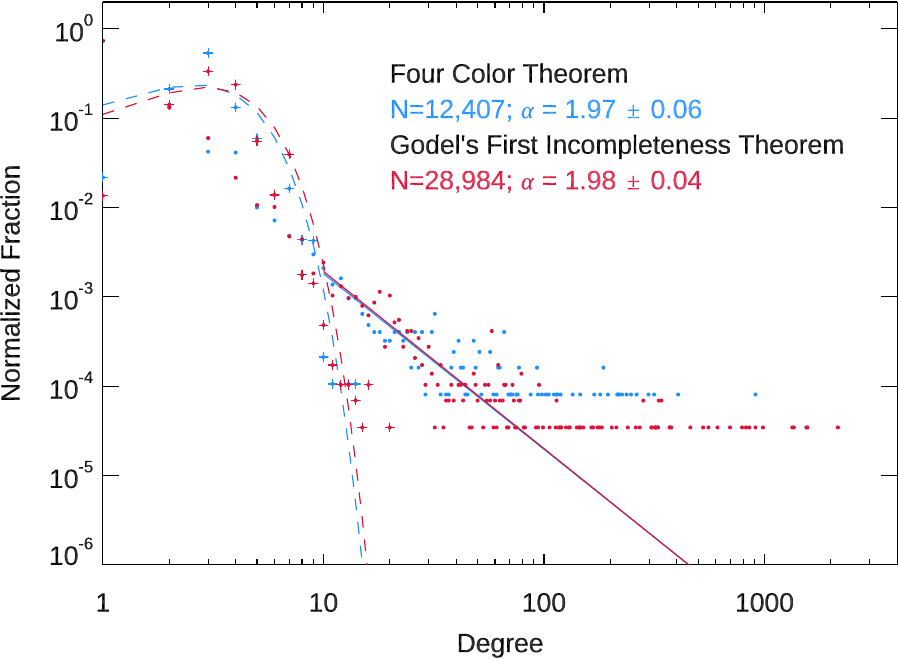}
    \caption{Distribution of in- ($+$/dashed fit) and out-degrees ($\cdot$/solid fit) for nodes in the automated proof of the Four Color Theorem and for G\"{o}del's First Incompleteness Theorem. While any node depends on a small number of others, following an Poisson distribution, the usage of a node in further claims follows a heavy-tailed distribution, with power-law index $\alpha$ around two.}
    \label{dist}
\end{figure}

\begin{table}[]
    \centering
    \begin{tabular}{r|l|l|l|l}
Theorem & Nodes & $\alpha$ & $f_2$ & $\Delta\mathcal{L}_1$ \\ \hline
{\bf Machine-Aided} & & & & \\
Euclid's \emph{Elements} &  174,597 & $2.14 \pm 0.02$ & $0.985$ & $+11$  \\ 
1st G{\"o}del Incompleteness &  28,984 & $1.98 \pm 0.04$ & $\star$ & $+13$  \\ 
Bertrand's Ballot &  24,137 & $2.22 \pm 0.05$ & $0.998$ & $+13$  \\ 
Polyhedron Formula &  23,750 & $2.22 \pm 0.04$ & $0.988$ & $+13$  \\ 
Euler's FLiT &  22,444 & $2.24 \pm 0.05$ & $0.998$ & $+11$  \\ 
Bertrand's Postulate &  22,434 & $2.09 \pm 0.04$ & $\star$ & $+15$  \\ 
F.T. Algebra &  20,431 & $2.29 \pm 0.05$ & $\star$ & $+14$  \\ 
Subsets of a Set &  20,205 & $2.13 \pm 0.05$ & $\star$ & $+9$  \\ 
Pythagorean Theorem &  18,230 & $1.93 \pm 0.04$ & $\star$ & $+19$  \\ 
Desargues's Theorem &  18,213 & $1.99 \pm 0.04$ & $0.987$ & $+18$  \\ 
Taylor's Theorem &  17,809 & $2.20 \pm 0.05$ & $\star$ & $+13$  \\ 
Heron's Formula &  17,487 & $2.04 \pm 0.04$ & $\star$ & $+17$  \\ 
F.T. Calculus &  16,845 & $2.24 \pm 0.05$ & $\star$ & $+13$  \\ 
Binomial Theorem &  16,314 & $2.11 \pm 0.05$ & $0.998$ & $+16$  \\ 
Geometric Series &  16,160 & $2.16 \pm 0.05$ & $\star$ & $+14$  \\ 
Wilson's Theorem &  16,120 & $2.09 \pm 0.05$ & $\star$ & $+12$  \\ 
Sylow's Theorem &  15,942 & $2.16 \pm 0.05$ & $\star$ & $+12$  \\ 
Ceva's Theorem &  15,279 & $2.11 \pm 0.05$ & $0.99$ & $+17$  \\ 
Bezout's Theorem &  14,909 & $2.23 \pm 0.06$ & $0.998$ & $+15$  \\ 
Reals Uncountable &  14,574 & $2.31 \pm 0.05$ & $0.996$ & $+14$  \\ 
Int.\ Value Theorem &  14,467 & $2.09 \pm 0.06$ & $0.998$ & $+12$  \\ 
Quadratic Reciprocity &  14,397 & $2.07 \pm 0.06$ & $0.998$ & $+13$  \\ 
Triangle Inequality &  13,657 & $2.14 \pm 0.06$ & $0.998$ & $+15$  \\ 
Leibniz $\pi$ &  13,619 & $1.91 \pm 0.05$ & $0.986$ & $+16$  \\ 
Pythagorean Triples &  13,254 & $2.19 \pm 0.06$ & $\star$ & $+14$  \\ 
Rationals Denumerable &  13,108 & $2.22 \pm 0.06$ & $0.998$ & $+13$  \\ 
Isosceles Triangle &  13,055 & $2.11 \pm 0.05$ & $0.995$ & $+15$  \\ 
Div 3 Rule &  13,037 & $2.22 \pm 0.06$ & $0.998$ & $+14$  \\ 
Inclusion-Exclusion &  12,886 & $2.15 \pm 0.06$ & $\star$ & $+16$  \\ 
Cauchy-Schwarz &  12,647 & $2.13 \pm 0.05$ & $\star$ & $+14$  \\ 
Four Color Theorem &  12,407 & $1.97 \pm 0.06$ & $0.989$ & $+10$  \\ 
Factor \& Remainder &  11,815 & $2.16 \pm 0.05$ & $0.997$ & $+13$  \\ 
Birthday Paradox &  11,692 & $2.10 \pm 0.06$ & $\star$ & $+14$  \\ 
Liouville's Theorem &  11,645 & $2.31 \pm 0.06$ & $\star$ & $+13$  \\ 
Cayley-Hamilton &  11,407 & $2.09 \pm 0.05$ & $\star$ & $+15$  \\ 
F.T. Arithmetic &  11,362 & $2.13 \pm 0.06$ & $\star$ & $+12$  \\ 
Cubic Solution &  11,271 & $1.88 \pm 0.05$ & $0.986$ & $+17$  \\ 
GCD Algorithm &  10,792 & $2.26 \pm 0.07$ & $0.932$ & $+14$  \\ 
Cramer's Rule &  10,613 & $2.09 \pm 0.05$  & $\star$ & $+15$  \\ 
Subgroup Order &  10,583 & $2.23 \pm 0.06$ & $\star$ & $+14$  \\ 
Mean Value Theorem &  10,168 & $2.20 \pm 0.07$  & $\star$ & $+13$  \\ 
Ramsey's Theorem &  7,747 & $2.37 \pm 0.09$& $0.997$ & $+13$  \\ 
Schroeder-Bernstein &  1,331 & $2.21 \pm 0.19$  & $0.987$ & $+14$  \\ 
Triangle Angles &  739 & $2.38 \pm 0.19$ & $0.99$ & $+11$  \\ 
Powerset Theorem &  282 & $2.41 \pm 0.32$ & $0.992$ & $+10$  \\ 
Prime Squares &  250 & $2.25 \pm 0.29$ & $0.983$ & $+10$  \\ 
Pascal's Hexagon &  150 & $2.32 \pm 0.34$  & $0.992$ & $+9$  \\ 
Induction Principle &  40 & ---  & $0.955$ & n.d.  \\ \hline
{\bf Human} & & & & \\
Euclid's \emph{Elements} &  475 & $1.97 \pm 0.07$ & $\star$ & $+21$  \\ 
Apollonius's \emph{Conics} &  446 & $2.28 \pm 0.11$ & $\star$ & $+12$  \\ 
Herstein \emph{Topics in Algebra} & 280 & $2.36 \pm 0.10$ & $0.957$ & $+5$ \\
Wiles's FLT &  142 & $3.39 \pm 0.72$ & $0.941$ & $+5$  \\ 
Orlick \& Strauch & 61 & $2.14\pm0.16$ & $0.951$ & $+6$ \\
\end{tabular}
    \caption{The statistics of dependence and implication in machine- and human-proved theorems (F.T.: ``Fundamental Theorem''; FLiT: ``Fermat's Little Theorem''). Both are characterized by high levels of modularity, and a long, power-law tail associated with assembly-and-tinkering construction. Over the entire dataset, machine proofs have a $\alpha$ equal to $2.15 \pm 0.13$. $f_2$: average degree of belief in final theorem at one-step error rate of $10^{-2}$; $\star$ indicates $f_2>0.999$. $\Delta\mathcal{L}_1$: log-likelihood penalty to within-module flip at $\beta$ equal to unity. Networks are truncated to the first depth expansion that produces more than 10,000 nodes, where possible otherwise to maximum depth.}
    \label{alpha_list}
\end{table}
Fig.~\ref{graphs} presents four examples of the proof networks we use in this analysis, with modules identified by the Girvan-Newman algorithm coloured, and with the size of a node indicating out-degree, or how many nodes depend upon it. By construction, the modules contain claims that are tightly linked together; in the case of the machine-aided proofs, for example, they contain claims that re-use some of the same intermediate variables, axioms, and prior results. Because there are far fewer nodes in the human-coded case, it is easier to directly translate modules into conceptual units. By inspection, for example, the light blue module for Euclid's \emph{Elements} is primarily composed of the incommensurability results of Book X, while the nearby light green cluster contains results on solid figures including the method of exhaustion in Book XII. The less tightly linked orange cluster contains a system of connected results in Book VII, VIII, and IX in number theory, while the smaller red cluster contains the small group of results in Book IX concerning the properties of even and odd numbers.

A key question for these networks concerns the extent to which some nodes are repeatedly re-used by a wide variety of others. One way to look more closely is Fig.~\ref{dist}, which shows the in- and out-degree distributions for the machine proof of the Four Color Theorem and G\"{o}del's First Incompleteness Theorem. While in-degree (the number of prior nodes a particular claim depends on) is Poisson-distributed, the out-degree (the number of nodes that use that claim) has a heavy-tailed distribution, with a fraction of nodes having influence hundreds of times larger than average. 

For readers unfamiliar with a degree-distribution plot like Fig.~\ref{dist}, consider first the blue $+$ symbols; these count the fraction of claims in the Four Color Theorem that have (going from left to right) one, two, and so on, dependencies (the ``in-degree''). For example, slightly more than 10\% of the nodes in the Four Color Theorem have two dependencies. The overall distribution is consistent with a Poisson distribution (blue dashed line), with the majority of nodes having three dependencies. By contrast, consider the blue $\cdot$ symbols, which count the fraction of claims with different numbers of things that depend on them, the ``out-degree''. Again, the majority of nodes have only a small number of things that depend on them, and (for example) less than 1\% have more than ten nodes that depend on them. However, a significant difference emerges now when looking at very high out-degree. In contrast to the Poisson distribution associated with the in-degree, the out-degree has a surplus of nodes that have very high out-degree. A similar ``heavy-tailed'' pattern appears for G\"{o}del's First Incompleteness Theorem, and indeed for all of the theorems, human and machine, seen in the data. 

This heavy-tailed distribution follows a power-law, with the probability of a node having degree $d$ given by
\begin{equation*}
    P(d) \propto d^{-\alpha}.
\end{equation*}
Across our sample of both machine and human proofs, these $\alpha$ values cluster tightly around two (Table~\ref{alpha_list}; fit using the methods of \citeA{clauset2009power}). 

This pattern, of both an exponential distribution for in-degree, and the particular value of $\alpha$ for the out-degree power-law tail, is a characteristic sign of the generative assembly model of \citeA{redner} and \citeA{redner2}. The network construction process of that model has two steps: first, a new node chooses some number of nodes to depend upon; second, from that set of chosen nodes, it chooses to link to some of their dependencies in a probabilistic fashion. It is found in both cultural and biological systems governed by successive accretion of links in a distinctive pattern associated with opportunistic ``tinkering and reuse''~\cite{sole}. 
``Tinkering and reuse'' means that a node is not simply used multiple times in a proof, but that parts of its precursors are used in new ways as well (the term ``tinkering'' comes from how this effect works in engineered and biological systems; here it makes more sense to talk in terms of ``repurposing''). 

Taking the example of Fig.~\ref{demo_figure}, imagine that claim $C$ was used twice in a larger proof, to establish, say, claims $C^\prime$ and $C^{\prime\prime}$. The tinkering and reuse process would make it more likely that $C^\prime$ would simultaneously rely on both $C$ and one of $C$'s precursors (say, $u$); meanwhile, $C^{\prime\prime}$ might make use of a different part of $C$, say $q$. While $u$ and $q$ might have been originally understood in terms of how the establish $C$, the construction process puts them to new uses associated with $C$'s ultimate task.

The general version of this process leads to a series of topological properties, including, characteristically, the values of $\alpha$ seen in Table~\ref{alpha_list}, as well as high levels of modularity consistent with the firewall effect discussed below. It provides one explanation for the unusual statistical properties of our networks, that fits with a natural generative process.

\subsection{Epistemic Phase Transitions and Modular Firewalls}

As described in above, we look for epistemic phase transitions as a function of both deductive and abductive implication strength: the degree to which the truth of a claim is coupled to the truth of either a claim it depends on, or a claim that it implies. We parameterize these by two terms, $\beta_{\mathrm{dep}}$ and $\beta_{\mathrm{imp}}$, for the two pairwise effects of truth (or falsehood). 

On the abductive side, $\exp{(2\beta_{\mathrm{imp}})}$ is the multiplicative factor by which a correct implication makes the node more likely to be true; on the deductive side, $\exp{(2\beta_{\mathrm{dep}})}$, the multiplicative factor by which a correct deduction makes the node more likely to be true. Taking (for simplicity) a symmetric error-making model, where the probability of incorrectly drawing a false conclusion from a true premise is the same as drawing a true conclusion from a false premise (and respectively for the abductive case), $\beta$ implies an error rate $\epsilon$ of
\begin{equation*}
    \epsilon = \frac{1}{1+e^{2\beta}},
\end{equation*}
which corresponds to the error rate of Hume's original paradox.

\begin{figure}[h!]
    \centering
    \begin{tabular}{c}
    (a) \\
    \cropped{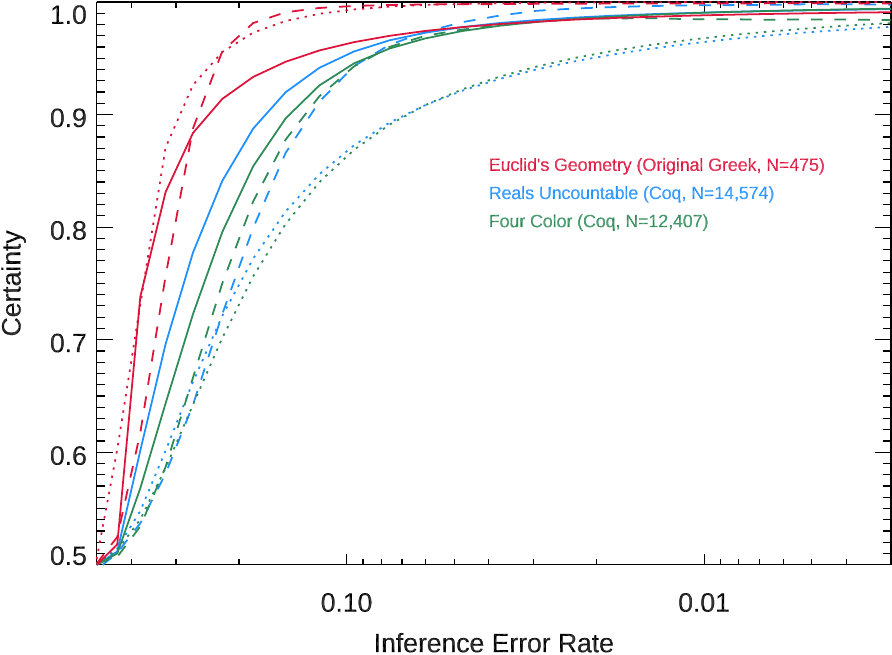} \\
    (b) \\
    \cropped{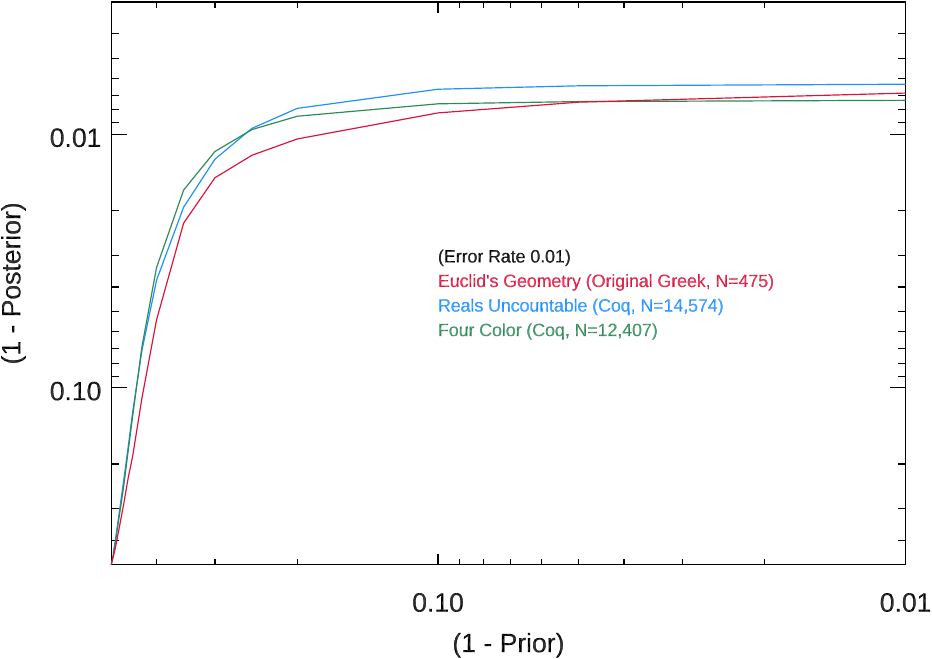}
    \end{tabular}
    \caption{Top: Epistemic phase transitions in three theorems: Cantor's theorem on the uncountability of the Reals, the Four Color Theorem, and Theorem IX.36 (the form of perfect numbers) in Euclid's Elements. Accuracy increases (and error rates decrease) going to the right. Solid lines indicate average degree of belief over all steps of the proof; dashed lines, in the theorem itself; dotted lines, in the axioms. Bottom: the relationship between prior and posterior, after equilibrating to the heuristic model, at an inference error rate of $0.01$. This error rate puts all three proofs past the phase transition point, and this means that even weak priors lead to near-unity degrees of belief. Remnant uncertainty at priors close to $0.5$ is due in part to frustrated freeze-in, \emph{i.e.}, where modules fall separately into all-true or all-false states.}
    \label{ept}
\end{figure}
Fig.~\ref{ept}a shows epistemic phase transitions in action for three proofs; Cantor's theorem on the uncountability of the Reals (Coq, $N=14,574$), the Four Color Theorem (Coq, $N=12,407$), and the (arbitrarily chosen) Proposition IX.36, the form of perfect numbers, of Euclid's \emph{Elements} with dependencies taken from the original Greek text ($N=475$). For simplicity in this case, we have set $\beta_{\mathrm{dep}}$ equal to $\beta_{\mathrm{imp}}$, and $p_\mathrm{prior}$ equal to $0.75$. 

We plot three quantities: the average degree of belief over all steps of the proof, the average degree of belief in the final theorem, and the average degree of belief in the axioms. For example, follow the red dashed line; this corresponds to the average degree of belief in the final claim of Theorem IX.36 in Euclid's \emph{Elements}. As the inference error rate goes from $0.5$ (\emph{i.e.}, maximal, equally likely to be true or false) down to $0.01$, the resulting degree of belief in the proof rises from $0.5$ (\emph{i.e.}, totally uncertain) to close to unity (total certainty). Crucially, despite the fact that this proof contains a large number of steps, and is thus subject to the Humean paradox, the overall certainty is much higher than expected; indeed, because of the existence of multiple paths, it is even better than what would be expected from the step-to-step inference error rate alone.

The three proofs in question show different certainty structures (for example, belief in the full proof lags that of both the theorem and axioms in the Euclidean case, while the reverse is true for the Four Color Theorem and the uncountability of the Reals), but share an overall pattern.  At cognitively-plausible error rates, the graph structure leads to a sharp transition where high levels of certainty emerge even when when error rates are at levels that would invalidate proofs made by deductive reasoning alone.

Column three of Table~\ref{alpha_list} shows $f_2$, the average degree of belief in the theorem (\emph{i.e.}, the terminal node) at a one-step error rate $\epsilon$ of $10^{-2}$. The majority reach near-unity levels of $f_2$ that exceed the one-step confidence. There are a few cases where this does not happen (\emph{e.g.}, Desargues's Theorem); this appears to be due, in part, to the presence of nodes just below the final theorem that have both few dependents and no other implications---these dangling assumptions participate only weakly in the larger network of justification.

Fig.~\ref{ept}b shows the effect of shifting $p_\mathrm{prior}$. To read this, consider, for example, the solid red line. If a person comes in to the proof with a prior that says any particular claim is 90\% likely to be true, then after following through the steps of the proof, and assuming an inference error rate of $0.01$, then, on average, the person will finish the proof with a 99\% degree of belief in any particular claim.

More formally, Fig.~\ref{ept}b shows that, past the critical point, even weak priors can lead to the transition to deductive certainty; in the physics-style language of phase transitions, this is the finite-size analogue of a divergence in the susceptibility. Failure in the case of weak priors is due to the emergence of domain walls, \emph{i.e.}, localized parts of the network that freeze into all-false or all-true states. This can lead either to (1) a cascade into the all-false condition driven by the small-number statistics of the fluctuations, or (2) a long-lived metastable state because interconnections are insufficiently strong to generate global consensus. Modular structure, which we discuss now, allows a practicing reasoner to escape the metastable state.

In particular, the resolution of our data allows us to characterize how modular structure creates topological ``firewalls'' where different parts of a proof decouple from each other. We compute the relative log-likelihood penalty to flip all the nodes in a module to the opposite truth value, versus, on the other, flipping the same number of nodes randomly chosen across the whole graph. We characterize this using $\Delta\mathcal{L}_1$, the log-likelihood penalty per nodes flipped, with the number of nodes set to ten and $\beta$ set to unity.

At high error rates, firewalls are fragile because there is little opportunity for order to propagate at any distance. However, as error decreases and order emerges, the distinct effects of within-module versus cross-module flips become apparent: the tighter connections between nodes within a module makes them easier to shift to the opposite state. As mathematicians increase their confidence their confidence in a proof, they find they can first achieve confidence in a particular module of the derivation even in the absence of strong beliefs about the truth in other places. This means that a modular proof strategy is easier than one that involves different parts of the system. Table~\ref{alpha_list} lists the $\Delta\mathcal{L}_1$ values, where a positive value indicates that within-module flips are more likely than cross-module ones. Values are around $+10$, corresponding to an overwhelming preference (at the $e^{+100}$ level) for within-, rather than cross-, module flips.

\subsection{The Abductive Paradox}

\begin{figure}[h!]
    \centering
    \begin{tabular}{c|c}
    \croppedsmall{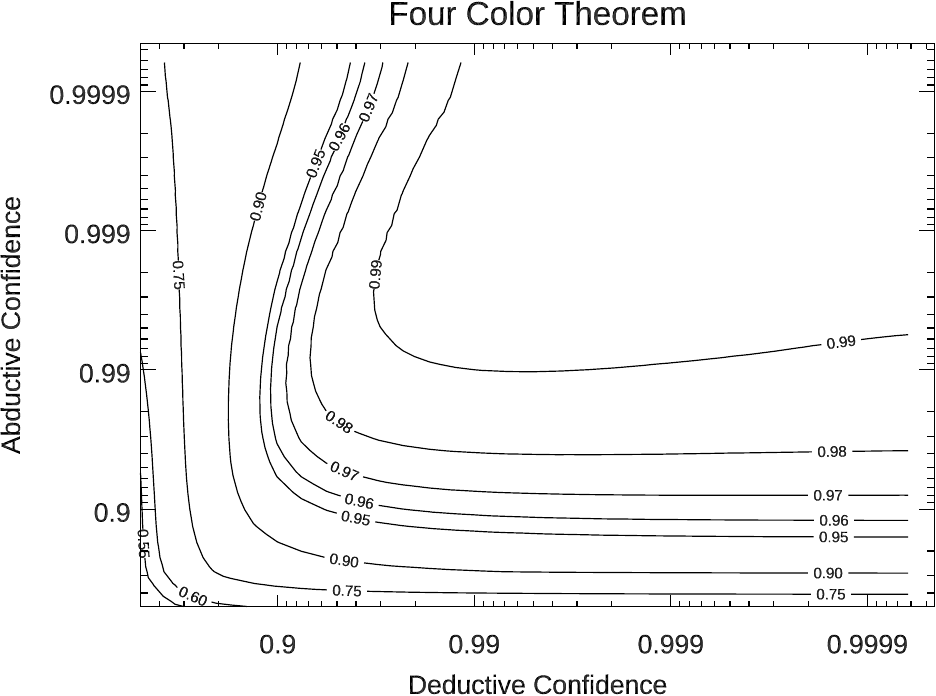} & \croppedsmall{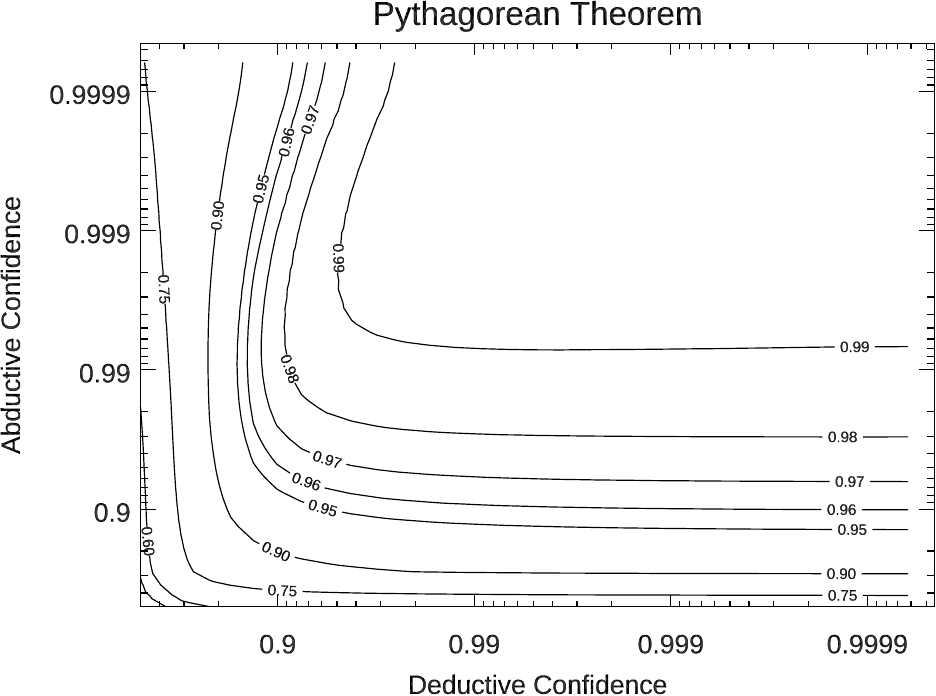} \\ \hline \\
    \croppedsmall{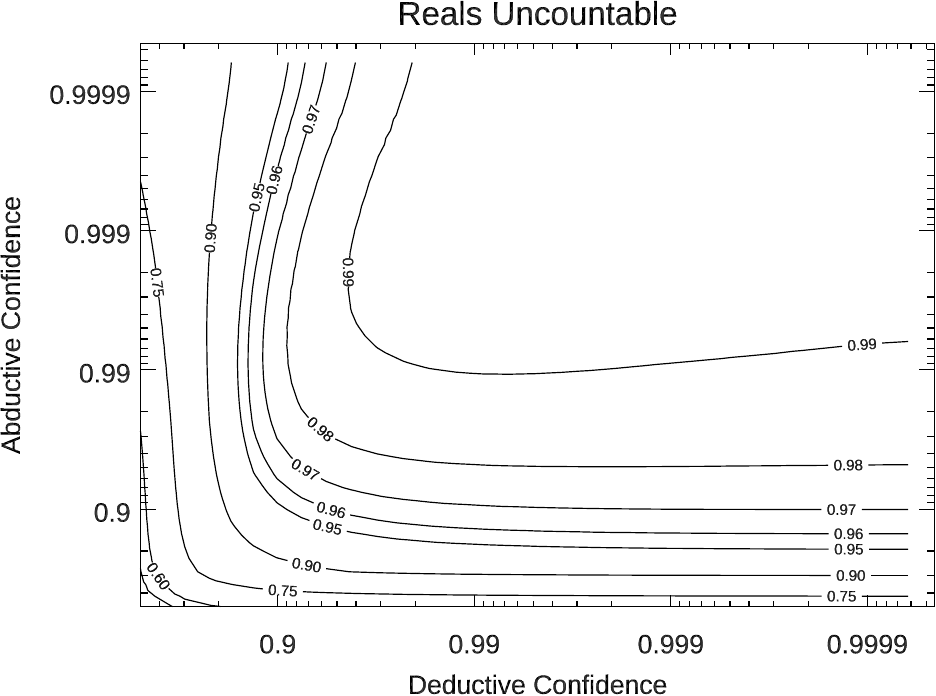} & \croppedsmall{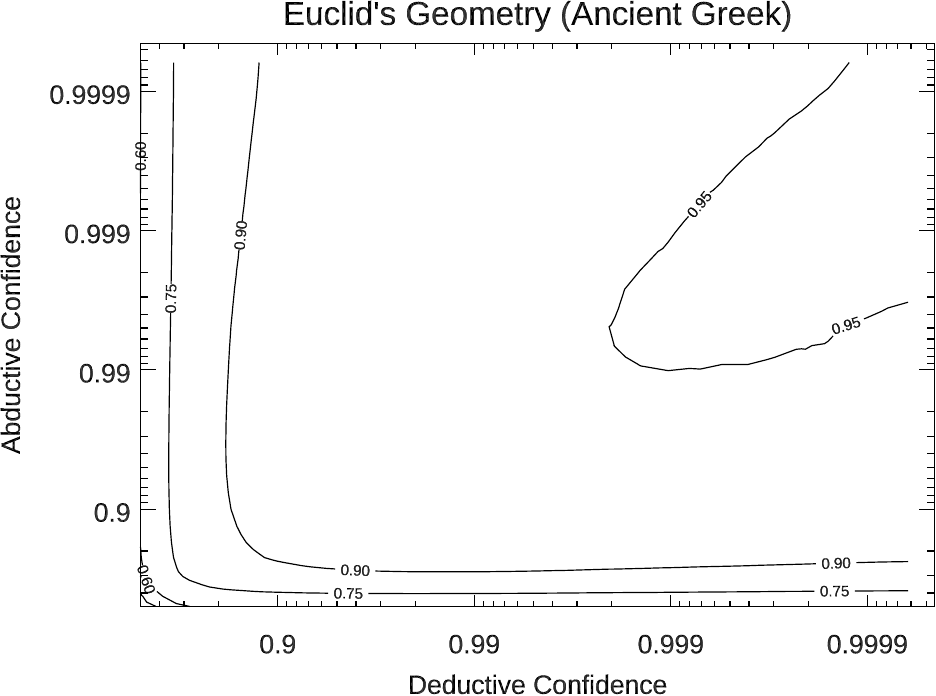} \\
    \end{tabular}
    \caption{Average degree of belief in four theorems and their preconditions, as a function of abductive and deductive confidence (confidence is defined as one minus error rate, so a confidence of $0.99$, for example, corresponds to an error rate of $0.01$). Network structure leads to levels of confidence far in excess of what can be expected on the basis of a linear derivation chain. For fixed deductive (abductive), but rising abductive (deductive) confidence, contours turn over, leading to an abductive paradox driven by an imbalance in the two modes of reasoning.}
    \label{contour}
\end{figure}
Fig.~\ref{contour} shows four examples of our third result. In each plot, we show how increasing deductive confidence (along the $x$-axis; equivalently, decreasing the error rate), or increasing abductive confidence rate (along the $y$-axis) alters the final level of certainty. For example, for the Four Color Theorem, a deductive error rate of $0.1$ (confidence $0.9$), and an abductive error rate of $0.1$ lead, together to an overall confidence in the proof a little higher than 90\%.

The certainty contours have an (at-first) unexpected property. Consider the Four Color Theorem, fix the deductive confidence to $0.9$, and vary the abductive error rate. As the abductive confidence increases from $0.5$ (at the bottom of the plot), the final certainty rises, as expected---the greater the confidence in one's abduction, the more confidence, it seems one has in the final result. As the abductive error rate drops below $0.01$, the reasoner reports confidence levels of 90\% and above. However, somewhere around an abductive confidence of $0.99$, the certainty starts to drop, as the contours ``turn over''. At an abductive confidence of $0.9999$, for example, the certainty is now \emph{below} 90\%---lower than it was when the agent had weaker confidence in their abductive abilities.

At a fixed level of deductive power, in other words, increasing abductive power can eventually lead to a degradation in the final degree of belief. A similar result obtains for the deductive case. In each case, for deductive (or abductive) certainty beyond the transition point, we see the contours of contrast belief turn over. A vertical (or horizontal) line drawn past the (rough) EPT point of $\epsilon$ equal to $0.1$ will eventually cross the contours going downwards in certainty. Contour plots for all the theorems discussed can be found in the Supplementary Information.

This is a general result. It is also a somewhat puzzling one, at first, and best clarified by example. Roughly speaking, a too-high confidence in the abductive pathway frustrates the bi-directionality necessary to achieve an EPT. In the extreme case, imagine a near-zero abductive error rate, \emph{i.e.}, an overpowering confidence in intution. If the mathematician comes to doubt the conclusion at all, the skepticism propagates downwards. Any beliefs about the reliability of the axioms, or the deductions made from them, that might overcome this skepticism have only weak influence further up.  The mathematician (sometimes) entertains intuitive, abductive doubts about the conclusion, he rates this abductive intuition much higher than his deductive powers, and the finishes by rejecting the proof completely. The claim is ``too crazy to be true''. Impervious to deductive arguments, he is certain a mistake has been made.

A similar effect takes place when the balance skews too far in favor of deductive reason, rather than abductive intuition. In this case, as the mathematician proceeds from the axioms up to the conclusions, she may eventually become concerned that a mistake has been made. At some point she entertains the possibility that claim currently to hand is false. Now, tragedy strikes, because everything further up, being based on false premises, can never establish the required claim.

Say, for example, the step in question combines a proof of $A$, and a proof that $A$ implies $B$, to derive $B$. If the hyper-deductive mathematician doubts the proof of $A$, the confidence she has in her deductive powers now works against the proof as a whole. She may have intuitions about the truth of all the consequences of the proof of $A$, but she doubts them too much to overcome the emerging deductive certainty that an error has been made and that everything that follows is flawed. A new proof might be constructed that establishes $B$ by other means. But the current proof, from her point of view, can not work.

\section{Discussion}

Our account of the emergence of mathematical belief depends on the use of paths that go both with and against the deductive grain to generate an epistemic phase transition. While this poses a challenge to the purely deductive model, the heuristics that underlie the EPT fit naturally with accounts that balance abduction and deduction, allow intuition to play a role in the status of a claim without coming to dominate deduction, and allow those intuitions to develop over time and in the course of examining a proof.

Consider, as an analogy, the use of computer code in a research project. Suppose I mostly believe some fact A, and I write a complex computer program to check and increase my confidence in that belief. If the program produces some output B that contradicts A, then I will likely first check the program itself for errors, a move that corresponds to doubting the axioms, or earlier stages of reasoning, in abductive fashion. Later, on reflection, I might realize that the output B is actually more plausible than A; this will now have the opposite effect, and my confidence in the earlier stages of the code will increase even if I do no further checks. Few theories of belief formation would rule out the analogous process in mathematical reasoning, which is also found in informal accounts by practitioners~\cite{villani}. A more elaborate reflection on the relationship between deduction and abduction, and the ways in which the two combine to produce a larger change in one's intuition about what is necessarily true, is provided by the mathematician and computer scientist Scott Aaronson.

\begin{quote}
    [A] step-by-step logical deduction tends to be seen as merely the vehicle
for dragging the reader or listener, kicking and screaming, toward a
gestalt switch in perspective---a switch that, once you've succeeded
in making it, makes the statement in question (and more besides)
totally obvious and natural and it couldn't ever have been otherwise.
The logical deduction is necessary, but the gestalt switch is the
point.  This, I think, explains the feeling of certainty that
mathematicians express once they've deeply internalized something---they're not multiplying out probabilities of an oversight in each
step, they're describing the scenery from a new vantage point that the
steps helped them reach (after which the specific steps could be
modified or discarded).~\cite{scottquote}
\end{quote}

Aaronson's remarks help us connect the restricted view provided by our focus on belief-formation, to the larger question of how proofs provide understanding and revise our mental models. If the certainty provided by the proof is \emph{part} of the understanding, our account, of how that belief formation can happen, helps us understand the nature of the new vantage point. 

Most obviously, our work suggests a key role for the presence of multiple, independent paths between claims. If it is the presence of these paths that makes the EPT possible, they may also play a role in structuring, or re-structuring, the mental model that the proof creates. In particular, our work suggests a role for how two claims (in our toy example from Fig.~\ref{demo_figure}, $t$ and $p$) are related to each other by a multiplicity of third terms (here, $q$ and $u$). Our results suggest that understanding a proof doesn't simply mean seeing that $t$ and $p$, say, ``are connected'', nor does it mean seeing that $t$ and $p$ are related by $q$; rather, it means having to hand a catalog of independent ways that $t$ and $p$ turn out to be related.

More speculatively, the fact that the multiple paths, and the required scaling for an EPT, can be achieved with a network structure associated with tinkering and reuse suggests that the method of construction may itself aid in the method of understanding-through-certainty. In particular, understanding may in part be driven not just by combining together different pieces of mathematics, but also through reusing and recombining some of their precursors. A proof can provide certainty in part because it shows how the precursor to an important theorem has other uses; conversely, a proof that simply combines pieces together without also reusing their internal components may not have the necessary structure to induce the revisionary understanding that comes from grasping a proof in its certainty.

Such a process fits with (but does not require) accounts of how proofs are actually made made. Lakatos' dialectic model, for example, presented in \emph{Proofs and Refutations}, emphasises tinkering by making proof an embedding of the truth of a conjecture into other areas of mathematics. The proof provides avenues for directed criticism, and both conjecture and proof are co-modified in response to the incremental introduction of local and global counterexamples. The structure of the final theorem is a product of this dynamic interplay. Mathematicians may tinker in ways that mimic far less rarified endeavours such as building an open source ecosystem~\cite{sole}. 

Another outcome of our work comes from our results on modular firewalls. They help us make sense of a common intuition among mathematicians that errors in a proof come in different forms, ``big'' and ``little'', or, in \citeA{robust}'s taxonomy, ``local'' and ``global''. Under a purely deductive model, of course, there is no distinction between a big error and a little one; any error vitiates the validity of the proof. If, however, we understand how proofs gather themselves into modular structures, then the distinction between local and global becomes much easier to understand. Work in the sociology of mathematics~\cite{grouptheory1} shows how the group theory community understood the difference in practice, with some mathematicians particularly prone to errors of local ``routine'' that, while irritating, could be fixed without affecting the overall structure of the proof. \citeA{habgood-coote_tanswell_2021} emphasizes how this distinction fits into the larger process of social knowledge production in mathematics: local errors may well be fixable, but only by those with expert knowledge in the particular locale, while global coherence is established by the community as a whole.

Indeed, the nature of the belief-formation process even suggests where the the two errors might be found. Local errors are more likely to occur at nodes in the middle of a densely-connected module, while the more serious global errors, that might require wholesale revision, occur at bottlenecks: nodes, or edges, where two densely connected clusters meet. In our toy example of Fig.~\ref{demo_figure}, with the addition of the $r$-node cluster, a bottleneck node would be node $r$; more generally, a network measure such as betweenness centrality~\cite{NEWMAN200539} is one computationally-tractable method for identifying where these global errors are more likely to appear. 

Attention to bottlenecks suggests a mechanism for how a collective, such as that examined by \citeA{habgood-coote_tanswell_2021}, can build confidence in a large proof even in the absence of what \citeA{tymoczko1979four} calls (individual-level) surveyability. We expect that the characteristic topologies of mathematical proofs in both the human and computer-aided cases likely bears on the broader question of how dialogic, and more broadly collective, processes bear on the establishment of social consensus in mathematical proof~\cite{novaes2020dialogical}, and on how informal fragments of mathematical reasoning can be assembled together to create values such as shareability and transferability used in \citeA{groundwork}'s recent fallibilist account of mathematical proof.

(The framework of our analysis, of course, is concerned with real-world practice, rather than any formal, normative justification. A theorem that contains a error in its logic may have little trouble deriving all sorts of abductively reasonable conclusions, and thereby lead mathematicians to believe, incorrectly but explosively, in its truth. It may be the case that the kinds of errors that invalidate proofs, in ways that can not be fixed, have distinct topological structures that prevent the emergence of an EPT. Such a determination requires a parallel database of invalid proofs.)


Our results also allow us to draw some conclusions about the nature of proofs produced without machine aid, and the ways in which mathematicians chose to draw attention to different steps in their construction. In the case of Wiles' proof of Fermat's Last Theorem, for example, we see significant deviations from the $\alpha$ equal to two power-law tail. This is due, in part, to a thinner network structure in which a text designed for human communication neglects to explicitly mention its reliance on common axioms or lemmas at every point they occur. This leads to a deficit of high-degree nodes whose absence frustrates an epistemic phase transition because they are no longer available as ``Grand Central Stations'' that increase the number of paths between points. Wiles' proof appears to be an outlier; the other human proofs in our data appear to have $\alpha$ close to two, indicating the presence of claims that are repetitively used and cited. In general, we find remarkable overlap in the properties of the machine and human aided proof networks, including the existence of an EPT, and the presence of modular firewalls.


A basic feature of our model is the idea that both deductive and abductive reasoning play a role in mathematical belief formation. In addition to accounts by practicing mathematicians, and philosophers of mathematics, a range of results in the psychology of mathematics education also support the idea. \citeA{harel1998students} provided an influential framework, ``proof schemes'', for understanding the different ways in which University students attempted to prove things mathematically. These proof schemes included both empirical and inductive pathways, as well as deductive ones. They also drew attention to how students refused to believe a proof until they could accumulate evidence from examples.  \citeA{TheRelationshipBetweenProofandCertaintyinMathematicalPractice} found similar patterns of behavior in a study of advanced doctoral students in mathematics, who were comfortable with the use of computational evidence as a way to increase their confidence in a conjecture, even in situations where they had had ample time to examine the logic of a (valid) proof. Other research into how students evaluate the validity of a proof, and what they consider to be a valid proof, shows that, for example, even in the presence of deductive errors, students will use other forms of evidence that proceed backwards from the claim~\cite{proving, KO201320}. 

Some of this literature focuses on how non-deductive justifications mislead students by, for example, blinding them to glaring errors in a proof's logic. From our point of view, this would count as an overvaluation of the abductive compared to the deductive, although with the added twist that the proof was designed to be invalid from the start. In other cases, there is a note of frustration with how students incorporate non-deductive arguments into the proofs themselves. This is, by mathematical norms, incorrect---but also, as mentioned in Section~\ref{cm}, a distinct problem from the one we consider, which is the use of abductive methods to evaluate a deductive argument.

That said, there is something strange about using abductive reasoning to form beliefs about a deductive argument. One is in the position of claiming, roughly speaking, that one is, on the basis of the proof, ``99\% certain that the claim is necessarily true''. The modal force of this claim is conceptually distinct from believing that the claim is true ``with degree of belief 99\%''. When mathematicians talk about verifying proofs, the former type of statement is quite common. For example, the mathematicians involved in peer-reviewing a recent proof in number theory wrote, in recommending its publication that, ``although we studied the arguments very thoroughly, we found it very difficult to spot even the smallest slip''~\cite{ny}. A proof may remove all doubt; the EPT makes it possible to be very, though not perfectly, confident that it has done so.

Beyond the abductive evaluation of deductive arguments, there is also something unusual about the bi-directional nature of the process. As a method of argument, without further qualification, it would be a case of begging the question: the plausibility of a conclusion leads one to think that the parts have been assembled correctly; the belief that the parts have been assembled correctly increases the confidence in the claim. Beneath this surface story, however, is the way that this back and forth movement might guide how mathematicians test their belief. A careful eye is guided to places, for example, where the results don't seem to make sense. One direction for future work is to use these models to determine where attention might be drawn, and to pick out key points in an argument where abductive, or deductive, justification will generate uncertainty.

It is not easy to understand how the deductive parts of any argument are justified. Common normative models, including Bayesianism~\cite{chater2008probabilistic}, fail because they assume logical omniscience, \emph{i.e.}, that the agent is aware of all the deductive consequences of their beliefs. If the agent is logically omniscient, however, then once it knows the axioms, it already knows the proof, and has no reason to read it. More speculative models do attempt to provide normative prescriptions in the absence of omniscience~(\emph{e.g.}, the ``logical induction'' of~\citeA{garrabrant2020logical}), but, in general, logical omniscience leads to the conclusion that any mathematical proof is always only begging the question of whether or not the axioms are true. In the final analysis, of course, a story of how one comes to belief a mathematical proof must be, according to most models, unusual. 

\section{Conclusion}

Ever since their invention in a cultural context associated with new forms of justification~\cite{axial}, mathematical proofs provide us with some of the most explicit examples of human reasoning we have available. They are an account of why we ought to believe something, that attempts to be immune to doubt.

Seen in this way, proofs are a test bed for justification practices more generally. Our results suggest that the confidence provided by an epistemic phase transition may also be present, in latent form, in many other kinds of claims we make in day-to-day life. As noted by \citeA{mercier2017enigma} in a cultural-evolutionary context, no piece of evidence---even one concerning the evidence of the senses---is transparent, and we can always be called upon to situate it in the context of a larger argument. This means that people expand claims about the physical world, or normative claims about how things ought to be, when they are asked for further justification. Those expansions, if constructed via a process similar to tinkering and reuse, ought to be able to support the same kinds of phenomena we establish here.

The combination of modularity, abduction, and deduction, as well as the underlying assembly mechanism of tinkering and reuse appears to have the power to generate significant levels of certainty even for complex and apparently fragile arguments. The networks we have studied here may be something that we create when called upon to justify our beliefs in non-mathematical claims as well.

Indeed, the justification of beliefs through reason is a basic task of the human species. Mathematical proofs provide an unusual example of what many cultures consider an ultimate standard. Our findings here suggest that underlying features of that justification can lead to firm beliefs, even when we understand ourselves to be fallible and limited beings.

\section{Acknowledgements}

We thank Jeremy Avigad, Kevin Zollman, Scott Aaronson, Colin Allen, Dennis Li, and Cait Lamberton for helpful discussions, our two anonymous referees for extensive comments and suggestions, and Kent Chang and Will Nalls for assistance with data entry. We acknowledge the support of the John Templeton Foundation and Jaan Tallinn via the Survival and Flourishing Fund.


\clearpage
\bibliographystyle{apacite}
\bibliography{preprint}

\end{document}